# *Optically Pumped Terahertz Amplitude Modulation in Type-II Ge QD/Si heterostructures grown via Molecular Beam Epitaxy*

Suprovat Ghosh[1], Abir Mukherjee[1], Sudarshan Singh[2], Samit K Ray[2], Ananjan Basu[1], Santanu Manna[3] and Samaresh Das[1,3,a)]

[1]*Centre for Applied Research in Electronics, Indian Institute of Technology Delhi, Delhi-110016, India*

[2]*Department of Physics, Indian Institute of Technology Kharagpur, Kharagpur-721302, India*

[3]*Department of Electrical Engineering, Indian Institute of Technology Delhi, Delhi-110016, India*

[a)] Author to whom correspondence should be addressed. Email: Samaresh.Das@care.iitd.ac.in

This article exploits group-IV germanium (Ge) quantum dots (QDs) on Silicon-on-Insulator (SOI) grown by molecular beam epitaxy (MBE) in order to explore its optical behaviour in the Terahertz (THz) regime. In this work, Ge QDs, pumped by an above bandgap near –infrared wavelength, exhibit THz amplitude modulation in the frequency range of 0.1-1.0 THz. The epitaxial Ge QDs outperform reference SOI (170 nm top Si) substrate in THz amplitude modulation owing to higher carrier generation in weakly confined dots compared to its bulk counterpart. This is further corroborated using theoretical model based on the non-equilibrium Green's function (NEGF) method. This model enables the calculation of photo carriers generated (PCG) and their confinement in the Ge QD region. Our model also reroutes the calculation from PCG to corresponding plasma frequency and hence to refractive index and THz photo-conductivity. Moreover, the photo-generated confined holes accumulation at the Ge QDs/Si interface is elevated after optical illumination, leading to a decreased THz photo-conductivity. This augmentation in THz photo-conductivity contributes to a significant enhancement of THz modulation depth ~77% at Ge QDs/Si interfaces compared to bare SOI at 0.1 THz.

## I. Introduction

Terahertz (THz) radiation has historically been an elusive frequency range (0.1–10.0 THz), lying between the infrared and microwave regions of the electromagnetic spectrum, with THz technology in its early stage. Nevertheless, the THz range has garnered significant attention due to its immense potential in various applications, including medical and biological sciences, security, imaging, non-destructive evaluation, and ultrafast computing. [1–4]. Initially, progress was impeded by the lack of powerful THz radiation sources, but over the past two decades, remarkable strides have been made in THz generation and detection. [5–7] These advancements have facilitated the emergence of THz time-domain spectroscopy for detailed material characterization [8] and non-invasive THz imaging for medical purposes. Despite progress, the lack of available components hampers effective THz wave manipulation, underscoring the need for further research; however, THz wave modulators show promise for information processing in THz wave communication and surveillance. Numerous techniques for developing THz wave modulator have been proposed, utilizing a range of materials such as graphene, transition metal chalcogenides (TMCs), [9–13] semiconductor-based meta-material, [14–16] phase transitional materials, [17,18] photonic crystal, [19] two-dimensional electron gas (2DEG) [20] and organic materials. [21] For instance, Wen et al. reported a graphene-on Ge (GOG) all-optical THz modulator with a modulation speed of 200 kHz and a modulation depth of 94%. [22] Mittendroff et al. demonstrated a complex waveguide-integrated THz modulator with a modulation depth of approximately 90%. [10] Researchers around the globe have also explored new TMD materials, such as $PtSe_2$, [23] as a promising material for THz modulation and reported a modulation depth of 32.7% at low-level optical pumping. While various THz modulator structures have been developed, their complexity limits practicality, especially for achieving high-speed modulation across a broad THz range. Quantum dots (QDs) show promise due to their confined quantum effects, fast carrier dynamics, tunable energy levels, and high carrier concentration, yet their potential in THz modulation remains largely untapped. This work introduces a reliable and effective THz amplitude modulation based on molecular beam epitaxy (MBE) grown epitaxial Ge QD/Si heterostructure. The modulation of the THz wave signal occurs through the generation of photo-induced carriers, which is controlled using an external continuous wave (CW) near-infrared laser. In this work, the THz amplitude modulation study of Ge QDs/Si heterostructure shows that it exhibits ~ 37% modulation depth at a pump power density of 5.3 W/cm$^2$, surpassing the performance of the bare SOI by 77 % at 0.1 THz. These heterostructure devices are easy to use, small in size, have a straightforward fabrication process, and are CMOS-compatible. These features make our design highly promising for wide range of applications, particularly in THz imaging and communication system. Furthermore, we have also demonstrated the theoretical background of such device performance based on non-equilibrium Green's function (NEGF) formalism.

## II. Material Growth and Characterisation

The MBE grown structure comprises a single layer of Ge QDs on SOI (MGQD-SOI) substrate, as illustrated in Fig. 1(a). The top Si thickness on the insulator was 170 nm. Growing Ge QDs involves using solid-source MBE via the Stranski-Krastanov (S-K) mechanism at a substrate temperature of 500˚C and a base pressure of ~ $2\times10^{-10}$ torr. Size of the grown Ge dots exhibit bimodal distributions [24–27], as can be seen from the atomic force microscopy (AFM) image using the Veeco Nanoscope-IV system, as shown in Fig. 1(b). The average height of the Ge QDs is 7–9 nm, with a corresponding base diameter ranging from 20–40 nm. The corresponding field emission scanning electron microscopy (FESEM) image is shown in the Fig. 1(c). Apart from these microscopic studies, the Raman spectra are acquired at room temperature using a T-64000 set-up from Jobin Yvon Horiba. An argon ion laser is used in this case, which is tuned to a 514.5 nm line. The set-up included an optical microscope that is equipped with a charged coupled device (CCD) to facilitate the acquisition of the spectra. In the Raman spectroscopy, Raman peaks corresponding to Ge-Ge and Si-Si vibrational bonds are observed at 300 $cm^{-1}$ and 520 $cm^{-1}$, respectively. This reveals that there is no intermixing of Si-Ge intermixing in the grown sample.

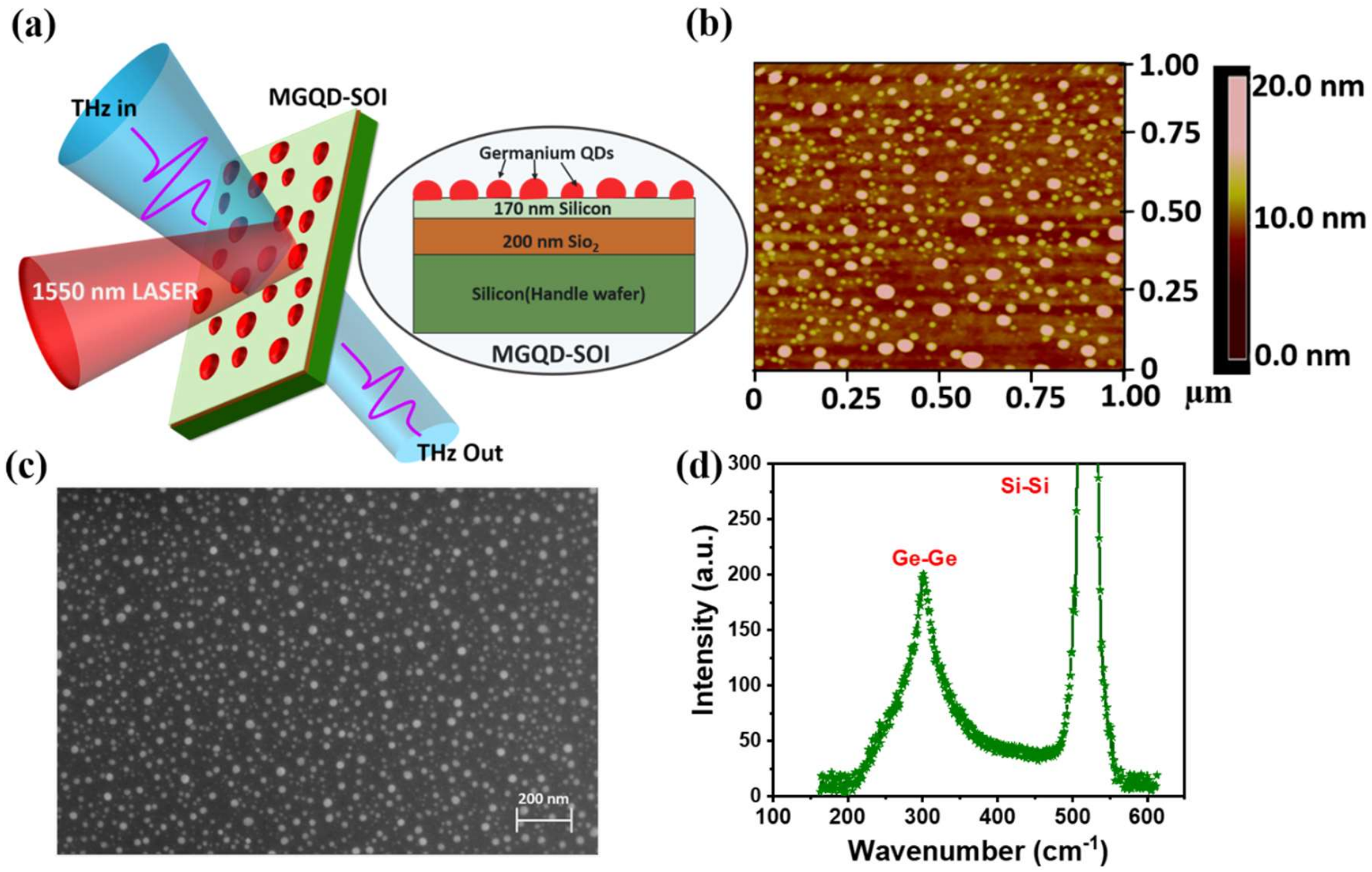

FIG. 1. (a) Schematic diagram depicting a THz amplitude modulation scheme based on MGQD-SOI, where the modulation is achieved via pumping of a near infrared laser (b) AFM image showing the surface topography of Ge QDs on a SOI substrate grown at about 500˚C (c) Top view FESEM image of the same sample and (d) Raman spectroscopy of the dot ensemble conducted in ambient environment.

## III. Terahertz Experimental Set Up

The experimental set-up for an optically controlled THz wave amplitude modulation is shown in Fig. 2. Both the THz wave and the pump laser are directed onto the MGQD-SOI device. The CW pump near-infrared laser spot on the device surface is 4 mm in diameter, while the THz beam spot size is ~3 mm. The THz spectrum is used as a reference without any device in the same environmental conditions. The measurements are performed using a wide-band CW frequency-domain THz equipment (TOPTICA's TERASCAN 1550) that covers a range of 0.1 THz to 1.0 THz. This system employs two distributed feedback (DFB) lasers of ~1533 nm and ~1538 nm that are combined by a coupler to generate an optical beat signal with a power of around 40 mW. While finely tuning the temperature of DFB lasers, it is possible to continuously vary the beat frequencies, generating a broad spectrum of THz frequencies ranging from 0.1 THz to 1.2 THz, with a frequency resolution of 50 MHz. In this set-up, an InGaAs photo mixer functions as a THz emitter, while a Schottky diode serves as a THz detector. An off-axis parabolic mirror is used to focus and direct the THz beam to the device and, subsequently, the transmitted beam to the receiver end. The induced photocurrent measured by the lock-in-amplifier at the Schottky diode is directly proportional to the THz intensity.

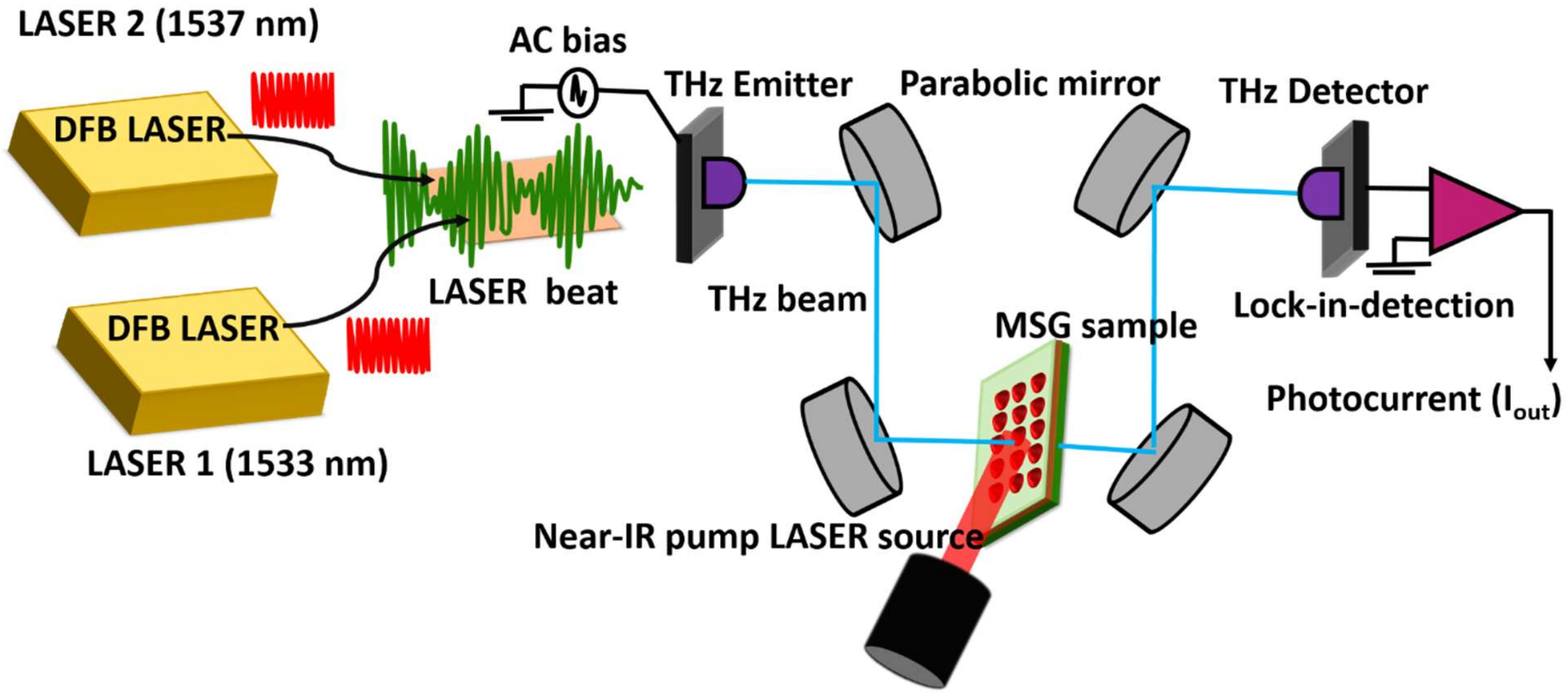

FIG. 2. Schematic diagram of the experimental set up for optically controlled THz modulation

## IV. Results and Discussions

Figure 3(a) shows the transmittance of the THz wave through the MGQD-SOI device at the different pump laser powers. In this case, an increment in laser power causes the decrement of THz transmission due to the increased photo-induced carrier density and, in turn, the change in the plasma frequency, which obeys the relation: $\omega_p = \sqrt{e^2 \rho / \varepsilon_0 m^*}$ , where $\rho$ is the photo-generated carrier density, which is a function of the pump power. In a later study, the potential effect of plasma frequency on the refractive index was calculated theoretically. The value of the THz transmittance $T\left(\omega_{THz}\right)$ of the devices is calculated at the frequency $\omega_{THz}$ using the relation:

$$T\left(\omega_{THz}\right)=\left[\frac{I_{sample}\left(\omega_{THz}\right)}{I_{reference}\left(\omega_{THz}\right)}\right]^2, \tag{1}$$

where $I_{sample}\left(\omega_{THz}\right)$ and $I_{reference}\left(\omega_{THz}\right)$ are the measured THz photocurrent with and without the device, respectively. Fig. 3(a) showcases the presence of Fabry-Perot oscillations, exhibiting distinct characteristics at different power levels. These oscillations are attributed to multiple reflections occurring at the surface of the sample as well as within the sample itself, which are caused by the refractive index variations. Further, the modulation depth (MD) is calculated as,

$$\text{MD}=\left|\frac{T_L-T_0}{T_0}\right|, \tag{2-a}$$

where $T_L$ and $T_0$ are the THz transmittance of the device with a pump power $P_L$ and without any power. Fig. 3(c) depicts the modulation depth of the MGQD-SOI device at various pump power densities. As the pumping power increases, there is an increase in the modulation depth. The highest modulation depth was recorded to be 37% at a frequency of 0.1 THz, corresponding to the illuminating power density of 5.3 W/cm$^2$. Fig. 3(d) illustrates the modulation depth of the modulator at various pump densities corresponding to the frequencies of 0.1 THz, 0.5 THz and 1 THz, respectively. Subsequently, we have calculated the relative percentage modulation depth change $RC_{QD}^{MD}$, with pump power density of 5.3 W/cm$^2$ (see Fig. 3(e)).

$$RC_{QD}^{MD}=\frac{MD_{QD}-MD_{SOI}}{MD_{SOI}}\times 100 \tag{2-b}$$

$MD_{QD}$ and $MD_{SOI}$ are the modulation depth of the MGQD-SOI and bare SOI, respectively. In addition, the dots also exhibit the dynamic modulation response probed at 0.1 THz under the periodic switching pump laser, as shown in Fig. 3(f). The laser beam undergoes modulation via a mechanical chopper at a frequency of 1 KHz. The dynamic response output of the modulator is detected by a Schottky photoconductive detector connected directly to an oscilloscope while varying light intensities. It can be noticed that under pump power density of 1.7 W/cm$^2$, modulation depth is lower than 5.3 W/cm$^2$owing to lower no. of carrier generation.

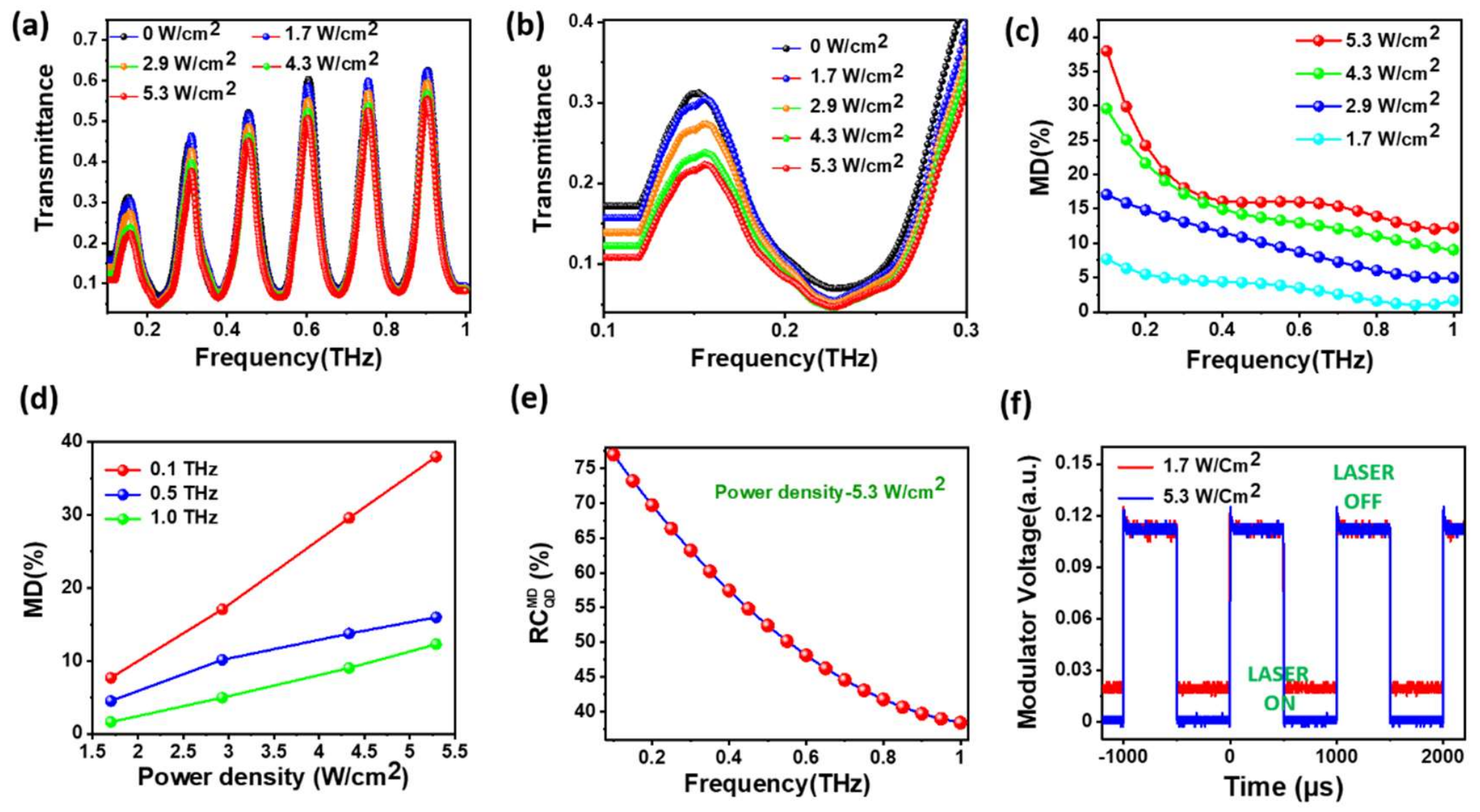

FIG. 3. (a) Transmittance plot of the MGQD-SOI sample in the frequency range of the 0.1 to 1.0 THz with different laser power density (b) Enlarged view of the transmittance plot in the frequency range of 0.1 to 0.3 THz range (c) Modulation depth (MD%) of the MGQD-SOI sample probed at 0.1 THz with different pump laser power density (d) Modulation depth (MD%) corresponding to the probe frequencies 0.1 THz, 0.5 THz and 1.0 THz with various pump power densities for same sample (e) Comparison of the percentage change in modulation depth induced by Ge QDs in the MGQD-SOI sample, relative to the bare SOI sample, across a broad frequency range of 0.1 THz to 1.0 THz. (f) The measured THz modulator voltage is at 0.1 THz frequency, considering different pumping power levels.

To deepen our understanding of the modulation mechanism stemming from the generation of photo-carriers within the Ge QD/Si heterostructure, we have employed the NEGF method. With the Ge QD/Si heterostructure exhibiting a Type-II band diagram (see Fig. 4 (a)), the valence band offset significantly surpasses the conduction band offset, measuring 0.4 eV and 0.05 eV, respectively. This notable discrepancy in energy levels implies that the photo-generated holes will be effectively confined within the Ge region, where as the electrons are inclined to migrate towards the Si layer. [28]

Capitalizing on this confinement effect makes it possible to manipulate the transmission of THz waves in the Ge QD/Si heterostructure. The accumulation of confined holes in the Ge region can be increased through optical pumping, thereby leading to a notable enhancement in the modulation depth for the MGQD-SOI device. This improved modulation depth holds great promise for various applications. Furthermore, an essential step involves finding the corresponding hole in-scattering function to accurately determine the concentration of confined photo-generated holes. This function is instrumental in incorporating the laser-induced disturbance within the QD. The expression for the hole in-scattering function is followed as: [29–31]

$$\left[\Sigma_{hole}\left(t,t'\right)\right]_{ij} = \sum_{l,k,\beta,\beta'} \left\{ \tau_{il}^{\beta} \left[ N_{\beta\beta'}^{ab-pht}\left(t,t'\right) n_{lk}^{c}\left(t,t'\right) \right]^{ISO} \tau_{kj}^{\beta'^{*}} \right\}, \tag{3}$$

The above equation can be transformed into energy domain by taking the Fourier transform. Here, the number of absorbed photons can be expressed as: $N_{\beta\beta'}^{ab-pht} = \left\{ \left( \frac{I_{\beta}}{E_{\beta}} \right) V_{ab} \delta_{\beta\beta'} \right\} \Big/ \tilde{\nu}$, where, $I_{\beta}$, $n^{c}$ and $V_{ab}$ correspond to the intensity of the laser source, correlation function of the filled states in the conduction band and the absorbing volume, respectively. Additionally, $\tau$ represents the interaction potential that governs the photo generation of electron-hole pairs within the device. Velocity of light within the material is $\tilde{\nu} = c/n^{\lambda}$, $n^{\lambda}$ being the refractive index of the Ge corresponding to the 1550 nm pump wavelength. In this case, the term $N^{ab-pht}\left(E_{\beta}\right)\, n^{c}\left(E-E_{\beta}\right)$ can be interpreted as the joint density of states. Moving forward, our objective is to determine the interaction matrix, which has been conceptualized as in the form of a scattering picture, as outlined $\tau = \left\langle f \middle| \widehat{V}_{e-\gamma} \middle| i \right\rangle$. The initial state (before scattering) is that of the virtual electron with hole effective mass with equal but opposite momentum, i.e., $\left| i \right\rangle = \left| -\vec{k}_i^{hole},\ m_h^* \right\rangle$ and the final state (after scattering) of real electron, $\left| f \right\rangle = \left| \vec{k}_f^{electron},\ m_e^* \right\rangle$. The carrier-photon interaction operator is: $\widehat{V}_{e-\gamma} = \frac{e}{2m^*}\left( \widehat{\vec{p}} \cdot \vec{A} + \vec{A} \cdot \widehat{\vec{p}} \right)$, where, $\widehat{\vec{p}}$ is the momentum operator and $\vec{A}$ is the vector potential of the incident photons. Thus, the light-matter interaction matrix can be written as: [32–34]

$$\left[ \tau^{\beta} \right] = \frac{e\hbar}{2}\left[ \left( \frac{\vec{k}^{hole}}{m_h^*} + \frac{\vec{k}^{electron}}{m_e^*} \right) . \vec{A}_0 \right] b_{\beta}^{Ph}, \tag{4}$$

where, $b_{\beta}^{Ph}$ is the field operator for the photon in $\beta$ mode and $\left| \vec{A}_0 \right| = \sqrt{\hbar^2 / 2\varepsilon E_{\beta} V_{ab}}$.

The modified Green's function for holes can be expressed as: [35,36]

$$G_h^M(E) = \left( [EI] - \left[ H_v^{ISO} \right] - \left[ \Sigma_{hole}(E) \right] \right)^{-1}, \tag{5}$$

where $H_v^{ISO}$ is the three-dimensional isolated Hamiltonian for the holes.

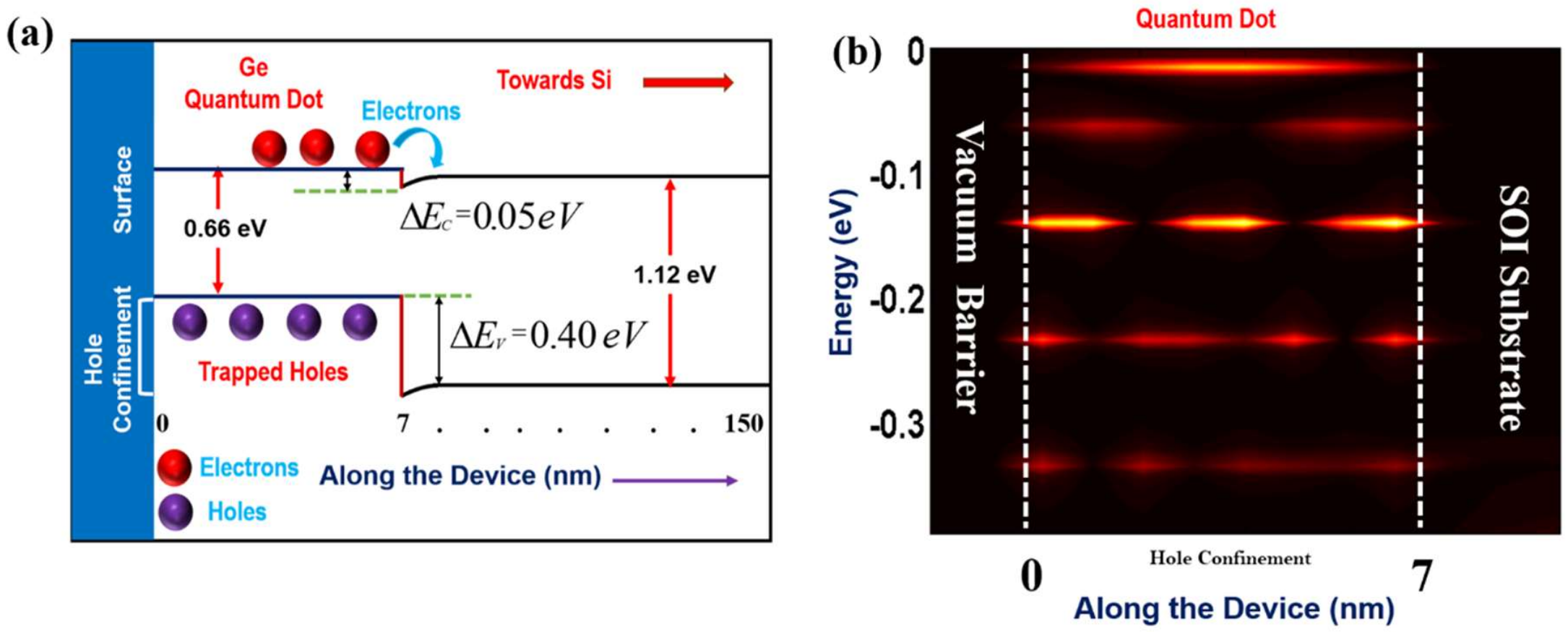

FIG. 4. (a) Schematic illustration of the energy band diagram showcasing the Ge QD and Si heterostructure. The diagram highlights the migration of photo-generated electrons towards Si while confining the holes within the QD. (b) Contour plot displaying the Local Density of States (LDOS), with an energy scale and the distance from the surface (represented as the vacuum barrier).

The total concentration of confined photo-generated hole concentration is calculated as follows: [32,37]

$$p=\frac{1}{\Delta}\int_{l_1}^{l_2}\left\{\int_{-\Delta\phi}^{0} D_{hole}^{Ph}(z,E)\,dE\right\}dz=\frac{1}{\Delta}\int_{l_1}^{l_2}\left\{\int_{-\Delta\phi}^{0}\left[G_h^M(E)\right]\left[\Sigma_{hole}(E)\right]\left[G_h^M(E)\right]^{\dagger}dE\right\}dz, \quad (6)$$

where, $D_{hole}^{Ph}$ and $\Delta$ are the local density of states of photo-generated holes and height of the QD, respectively. Total confined photo-generated hole concentration is evaluated by computing the integral in the position space for the QD region $(l_1, l_2 \in \mathrm{QD})$ and integral in the energy scale within the confining barrier, equivalent to valence band offset. Thus, the carrier density is $\rho = p / A_{\mathrm{QD}}$ where $A_{\mathrm{QD}}$ is the surface area of the QD. The calculation of the photo-generated hole density has been performed across a range of pump power and frequency conditions, as depicted in Fig. 5(a). The white dotted line represents the laser frequency chosen explicitly for the THz modulation experiment (MGQD-SOI device), corresponding to a photon energy of 0.80 eV or a wavelength of 1550 nm. Based on the graph, it is evident that the generation of photo carriers, particularly the concentration of holes, exhibits a substantial increase for wavelengths exceeding 1550 nm. The change in the dielectric constant for the Ge (QD region) in the THz regime is attributed to the presence of confined photo-generated holes, which can be described as follows [38]:

$$\Delta\left(\frac{\varepsilon}{\varepsilon_0}\right) = -\frac{{\omega_P}^2}{{\omega_{THz}}^2} = -\frac{e^2\rho}{{m_h}^*\varepsilon_0{\omega_{THz}}^2}, \tag{7}$$

Thus, the new modified refractive index is:

$$n_{New}^{THz} = n_{Dark}^{THz} - \frac{e^2\rho}{2{m_h}^*\varepsilon_0 n_{New}^{THz}{\omega_{THz}}^2}, \tag{8}$$

Where $n_{Dark}^{THz}$ is the refractive index of Ge without a pump laser. The modified refractive index of the QD region at 0.1 THz frequency for different pump photon energy and power is shown in Fig. 5(b). Equivalently, an external optical pumping results in effective doping and results in a significant change in the refractive index by generating the excess carriers. [39,40]

To obtain the THz photo conductivity, the following relation is used: [21,41]

$$\sigma(\omega_{THz}) = \left[\frac{4n(\omega_{THz})\exp\left[-i\left(n(\omega_{THz})-1\right)k_0 d\right]}{T(\omega_{THz})\left(1+n(\omega_{THz})\right)} - 1 - n(\omega_{THz})\right] \Big/ Z_0, \tag{9}$$

The Theoretically calculated transmission spectra and refractive index denoted as $T(\omega_{THz})$ and $n(\omega_{THz})$, are both obligatory to determine the THz photo conductivity. The change in the THz photo conductivity at the frequency range of 0.1–1.0 THz are substantial and are depicted in Fig. 5(c), as a function of laser power. Notably, our theoretical calculations of THz photoconductivity and refractive index did not account for Fabry-Pérot oscillations. Fig. 5(d) displays the modified refractive index at frequencies ranging from 0.1 to 1 THz while showing the impact of different pump power corresponding to the 1550 nm wavelength. Fig. 5(e) depicts a 3D plot illustrating the theoretically calculated modulation depth at various THz frequencies, alongside different lasing powers. We compared our experimental observations of the modulation depth to the theoretical results depicted in Figure 5(f) for lasing power density 5.3 W/cm$^2$, spanning frequency up to 0.3 THz. The discrepancy in modulation depth at higher THz frequencies is attributed to excluding the Fabry-Pérot effect in our theoretical analysis.

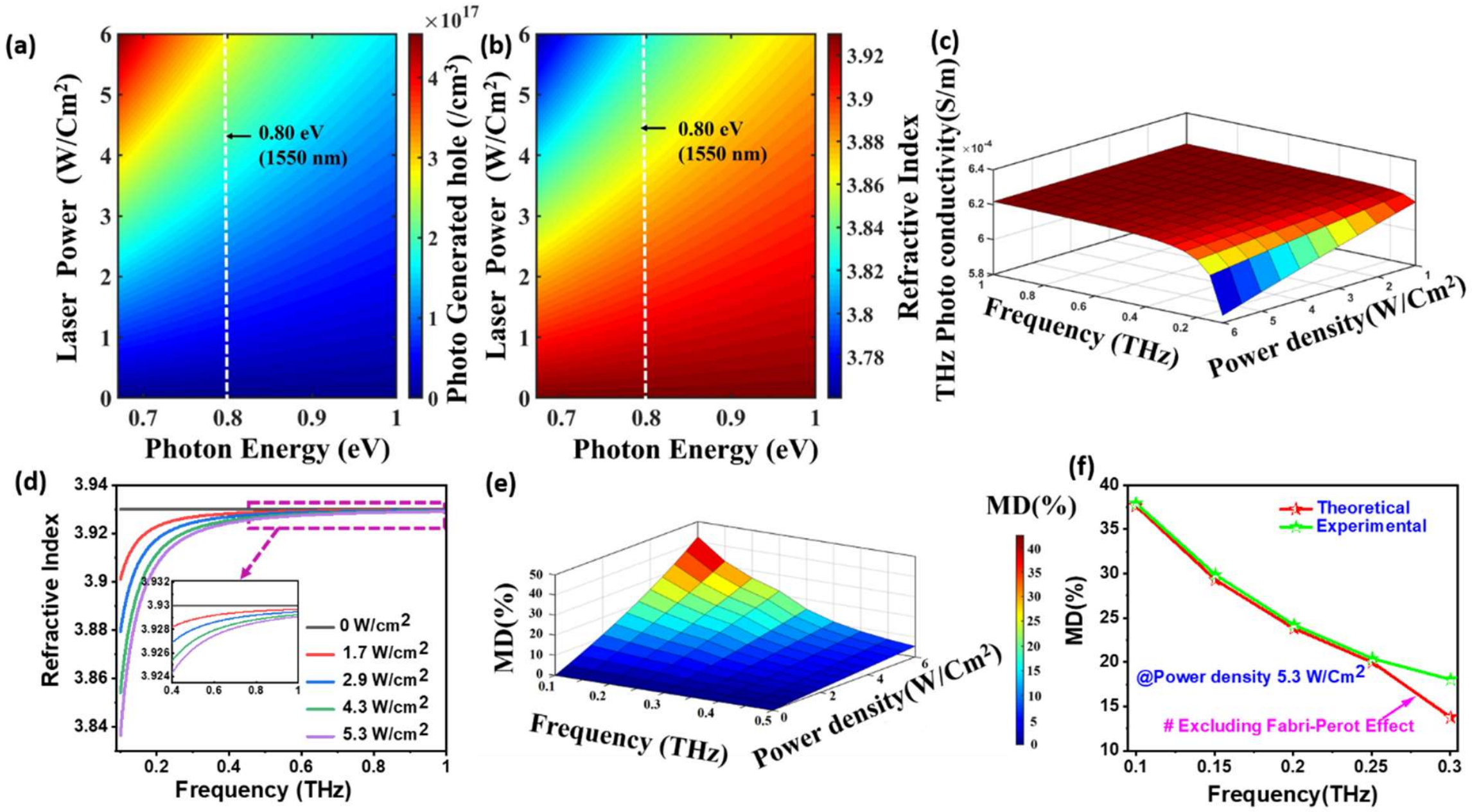

FIG. 5. (a) Contour plot illustrating the variation of confined photo-generated hole density as a function of pump photon energy, considering different pump power density values, (b) Corresponding refractive index at 0.1 THz frequency, demonstrating the influence of pump laser energy and power density variations. (c) THz Photo conductivity at different pump power density levels. (d) Refractive index at different THz frequencies ranging from 0.1 to 1.0 THz, using a near-infrared pump laser with varying power density. The inset offers an enlarged view of the THz range from 0.4 to 1.0 THz. (e) 3D plot of theoretically calculated modulation depth (MD%) MGQD-SOI sample against different laser power densities. (f) The modulation depth (MD%) of both experimental and theoretical results is compared at a lasing power density of 5.3 W/cm$^2$, spanning frequencies up to 0.3 THz.

## V. Conclusion

In summary, our experiment successfully demonstrates an optically controlled THz amplitude modulation using epitaxial Ge QDs grown on a SOI substrate. This Ge QD devices exhibit stable operation CW near-infrared laser with low pump power. Our results reveal a wide-spectrum modulation of THz transmission, spanning from 0.1 to 1.0 THz. Specifically, at 0.1 THz, the MGQD-SOI device showcased a significant 77% increase in modulation depth compared to the bare SOI. Theoretical calculations provided the impact of laser intensity and wavelength (energy) on the change in refractive index and THz photo conductivity, attributed to the generation of photo carriers within the Ge QD region. This also suggests that pump wavelength near the band gap can best impact the refractive index. Moreover, these Ge dots have several advantages, such as ease of fabrication and CMOS compatibility, which exhibit modulation in the THz regime. Green's function method begins by nabbing from a fundamental route to the THz application point of view, but also with possible future scopes of development for photonics application starting from exciton to polariton formation.

## ACKNOWLEDGMENTS

This work was supported partially by Science and Engineering Research Board (SERB) and Technology Development Program, Department of Science and Technology (DST), India with grant no CRG/2020/002262. S.K.R. acknowledges Meity sponsored NNETRa research grant at IIT Kharagpur. S.M. acknowledges support from IIT Delhi seed grant award. The authors would like to express their gratitude to Kritika Bhattacharya and Harmanpreet Kaur Sandhu for their valuable discussions.